\documentclass[aps,prc,letterpaper,11pt,twoside,tightenlines,nofootinbib,showpacs,preprint]{revtex4-1}
\usepackage{graphicx}
\usepackage[sort&compress]{natbib}
\usepackage{subfigure}
\usepackage{amsmath}
\usepackage{amsfonts}
\usepackage{cancel}

\begin{document}

\arraycolsep1.5pt

\newcommand{\Ima}{\textrm{Im}}
\newcommand{\Rea}{\textrm{Re}}
\newcommand{\mev}{\textrm{ MeV}}
\newcommand{\be}{\begin{equation}}
\newcommand{\ee}{\end{equation}}
\newcommand{\ba}{\begin{eqnarray}}
\newcommand{\ea}{\end{eqnarray}}
\newcommand{\gev}{\textrm{ GeV}}
\newcommand{\nn}{{\nonumber}}
\newcommand{\dtres}{d^{\hspace{0.1mm} 3}\hspace{-0.5mm}}

\title{Determination of the $\eta^\prime$-nucleus optical potential }
\author{M.~Nanova$^{1}$, V. Metag$^{1}$,~E.~Ya.~Paryev$^{2}$,~D.~Bayadilov$^{3,4}$,~B.~Bantes$^{5}$,~R.~Beck$^{3}$,~Y.~A.~Beloglazov$^{3,4}$,\\
~S.~B\"ose$^{3}$,~K.-T.~Brinkmann$^{1}$,~Th.~Challand$^{6}$,~V.~Crede$^{7}$,~T.~Dahlke$^3$,~F.~Dietz$^{1}$,\\~P.~Drexler$^{1}$,
~H.~Eberhardt$^{5}$,~D.~Elsner$^{5}$,~R.~Ewald$^{5}$,~K.~Fornet-Ponse$^{5}$,~S.~Friedrich$^{1}$,\\~F.~Frommberger$^5$,~Ch.~Funke$^{3}$,
~M.~Gottschall$^3$,~A.~Gridnev$^{3,4}$,~M.~Gr\"uner$^3$,~E.~Gutz$^{1,3}$,\\~Ch.~Hammann$^3$,~D.~Hammann$^{5}$,
~J.~Hannappel$^5$,~J.~Hartmann$^{3}$,~W.~Hillert$^{5}$,~P.~Hoffmeister$^3$, Ch.~Honisch$^3$, I.~Jaegle$^{6,a}$, 
D.~Kaiser$^3$, H.~Kalinowsky$^3$, S.~Kammer$^5$, I.~Keshelashvili$^6$,\\~V.~Kleber$^5$,~F.~Klein$^5$, E.~Klempt$^3$,~B.~Krusche$^6$, M.~Lang$^3$,~I.~V.~Lopatin$^{3,4}$,~Y.~Maghrbi$^6$,\\ K. Makonyi$^{1,b}$,~J.~M\"uller$^3$,~T.~Odenthal$^3$,~D.~Piontek$^{3}$,~S.~Schaepe$^3$,~Ch.~Schmidt$^3$, H.~Schmieden$^5$,~R.~Schmitz$^3$,~T.~Seifen$^3$,~A.~Thiel$^3$,~U.~Thoma$^3$,~H.~van~Pee$^3$,~D.~Walther$^3$,\\~Ch.~Wendel$^3$, U.~Wiedner$^8$,~A.~Wilson$^{7,3}$, A.~Winnebeck$^3$, and F.~Zenke$^3$\\
(The CBELSA/TAPS Collaboration)}
\affiliation {
{$^{1}$ II. Physikalisches Institut, Universit\"at Gie{\ss}en, Germany}\\
{$^{2}$ Institut of Nuclear Research, Russian Academy of Sciences, Moscow, Russia}\\
{$^{3}$Helmholtz-Institut f\"ur Strahlen- u. Kernphysik Universit\"at Bonn, Germany }\\
{$^{4}$Petersburg Nuclear Physics Institute, Gatchina, Russia}\\
{$^{5}$Physikalisches Institut, Universit\"at Bonn, Germany}\\
{$^{6}$Physikalisches Institut, Universit\"at Basel, Switzerland}\\
{$^{7}$Department of Physics, Florida State University, Tallahassee, FL, USA}\\
{$^{8}$Physikalisches Institut, Universit\"at Bochum, Germany}\\
{$^{a}$Current address: Hawaii University, USA}\\
{$^{b}$Current address: Stockholm University, Stockholm, Sweden}\\
}

\date{\today}
%
\begin{abstract}
The excitation function and momentum distribution of $\eta^\prime$ mesons have been measured in photon induced reactions on $^{12}{}$C in the energy range of 1250-2600 MeV. The experiment was performed with tagged photon beams from the ELSA electron accelerator using the Crystal Barrel and TAPS detectors. The data are compared to model calculations to extract information on the sign and magnitude of the real part of the $\eta^\prime$-nucleus potential. Within the model, the comparison indicates an attractive  potential of  -($37 \pm 10(stat)\pm10(syst)$) MeV depth at normal nuclear matter density. Since the modulus of this depth is larger than the modulus of the imaginary part of the $\eta^\prime$-nucleus potential of -($10\pm2.5$) MeV, determined by transparency ratio measurements, a search for resolved $\eta^\prime$-bound states appears promising.

 \end{abstract}
\maketitle

\section{Introduction}
\label{Intro}
Understanding the structure of low-lying hadrons is one of the challenging problems in the non-perturbative regime of quantum chromodynamics. The remarkably large mass of the $\eta^\prime $ meson compared to the masses of the other members of the pseudoscalar meson nonet is attributed to the $U_A(1)$ anomaly and to the explicit and dynamical breaking of chiral symmetry~\cite{Klimt,Jido}. One way to learn more about the interplay of these symmetry breaking effects is to study modifications of the $\eta^\prime$ mass in a strongly interacting environment where a partial restoration of chiral symmetry is expected. As a consequence of a reduction of the chiral condensate, a comparable drop in the U$_{A}(1)$ breaking part of the $\eta^\prime$ mass might be expected~\cite{Nagahiro,Kwon}. These predictions are, however, in conflict with earlier calculations within the Nambu-Jona-Lasinio-model which expect almost no change in the $\eta^\prime$ mass as a function of nuclear density~\cite{meissner}. It is obvious that these contradictory theoretical predictions call for an experimental clarification. \\
\par
A reduction of the $\eta^\prime$ mass by at least 200 MeV/$c^2$ in the collision zone of ultra-relativistic heavy-ion collisions was claimed in an analysis of  RHIC experiments~\cite{Csorgo}. On the other hand, based on the determination of the $\eta^\prime$ scattering length of $\vert a_{\eta^\prime}\vert \approx $ 0.1 fm in the $p p \rightarrow p p \eta^\prime$ reaction~\cite{Moskal}, a potential depth of only 8.7 MeV was predicted~\cite{Nagahiro_Oset}. A similar potential depth is expected in the quark meson coupling model~\cite{Bass}. In the present work, an attempt is made to extract the depth of the real part of the $\eta^\prime$-nucleus potential from the measurement of the excitation function and the momentum distribution of $\eta^\prime $ mesons in a photonuclear reaction. \\
\par
The in-medium properties of the $\eta^\prime$ meson are related to the $\eta^\prime$-nucleus optical potential which can be written as:
\begin{equation}
U_{\eta^\prime}(r) = V(r) + iW(r),
\end{equation}
where $V$ and $W$ denote the real and imaginary parts of the optical potential, respectively, and $r$- is the distance between the meson and the center of the nucleus. The $\eta^\prime$ in-medium mass shift $\Delta m(\rho_{0})$ at saturation density $\rho_{0}$ can be related to the strength of the real part \cite{Nagahiro1}
\begin{equation}
V(r) = \Delta m(\rho_{0})\cdot c^2\cdot \frac{\rho(r)}{\rho_{0}}.
\end{equation}
The imaginary part of the potential describes the meson absorption in the medium and is connected to the in-medium width $\Gamma_{0}$ of the meson at normal nuclear matter density by
\begin{equation}
W(r) = -\frac{1}{2}\Gamma_{0}\cdot \frac{\rho(r)}{\rho_{0}}.
\end{equation}

Experimentally, the imaginary part of the potential can be determined in a measurement of the in-medium width of the $\eta^\prime$ meson. As has been shown in~\cite{nanova}, the in-medium width of the $\eta^\prime $ meson can be extracted from the attenuation of the $\eta^\prime$ meson flux deduced from a measurement of the transparency ratio for a number of nuclei. For the $\eta^\prime$ meson, an in-medium width of 15-25 MeV at saturation density has been reported for an average recoil momentum $p_{\eta^\prime}$ = 1.05\ GeV/c~\cite{nanova}.
Taking into account Eq.(3), the imaginary part of the optical potential at this density is thus determined to be $W(\rho_{0})=-(10.0\pm2.5)$ MeV.\\
 Information on the real part of the meson-nucleus potential and thereby on the in-medium mass can be extracted from a measurement of the excitation function and momentum distribution of a meson as discussed in~\cite{Weil}. A downward shift of the meson mass would lower the threshold for meson photoproduction. Due to the enlarged phase space, the production cross section for a given incident beam energy will increase as compared to a scenario without mass shift. Furthermore, mesons produced in a nuclear reaction leave the nuclear medium with their free mass. In case of an in-medium mass drop, this mass difference has to be compensated at the expense of their kinetic energy.  As demonstrated in GiBUU  transport-model calculations~\cite{Weil}, this leads to a downward shift in the momentum distribution as compared to a scenario without mass shift. A mass shift can thus be indirectly inferred from a measurement of the excitation function and/or the momentum distribution of the meson. 
For the $\eta^\prime$ meson, this idea has independently been pursued on a quantitative level by Paryev~\cite{Paryev}.

\section{Experiment}
The experiment was performed at the ELSA electron accelerator facility~\cite{Husmann_Schwille,Hillert} at the University of Bonn, using the Crystal Barrel (CB)~\cite{Aker} and Two Arm Photon Spectrometer (TAPS)~\cite{Novotny,Gabler} detector system which provides an almost complete coverage of the full solid angle for decay photons of mesons produced in the photon induced reaction. Data on $\eta^\prime$ photoproduction were taken during the beam time period in January 2009 within 470~h effective running time.  Tagged photons of energies 0.7 - 3.1 GeV were produced via bremsstrahlung from an electron beam of 3.2 GeV. The energy of outgoing electrons and, thus, that of the incident photons was known with an accuracy of 3-18 MeV for the given energy regime. The total rate in the tagging system was 10\ MHz. A detailed description of the tagger setup can be found in an earlier publication~\cite{Elsner}. \\
\begin{figure}
 \resizebox{1.\textwidth}{!}
  {
   \includegraphics[height=0.9\textheight]{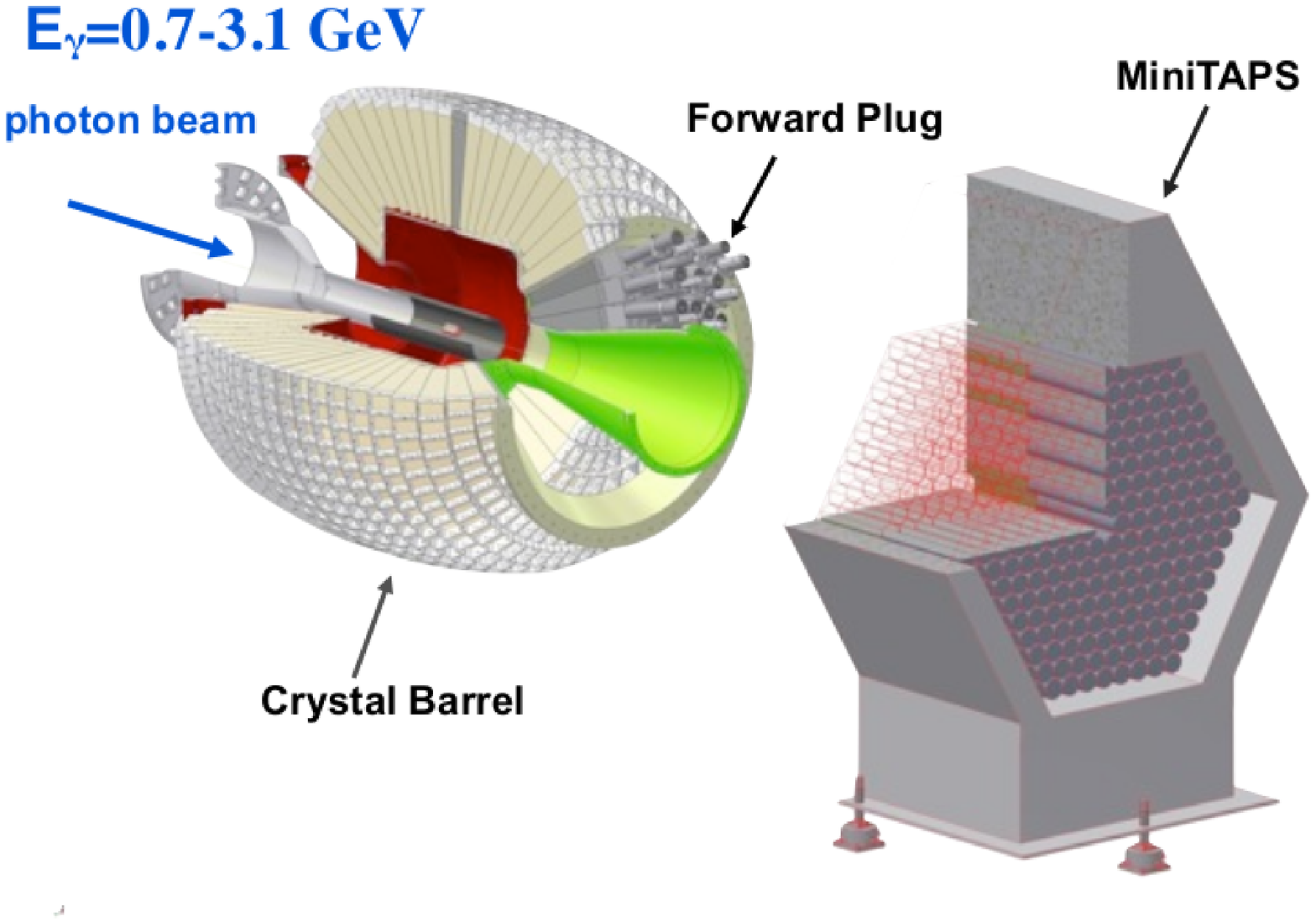}  \space  \  \includegraphics[height=0.9\textheight]{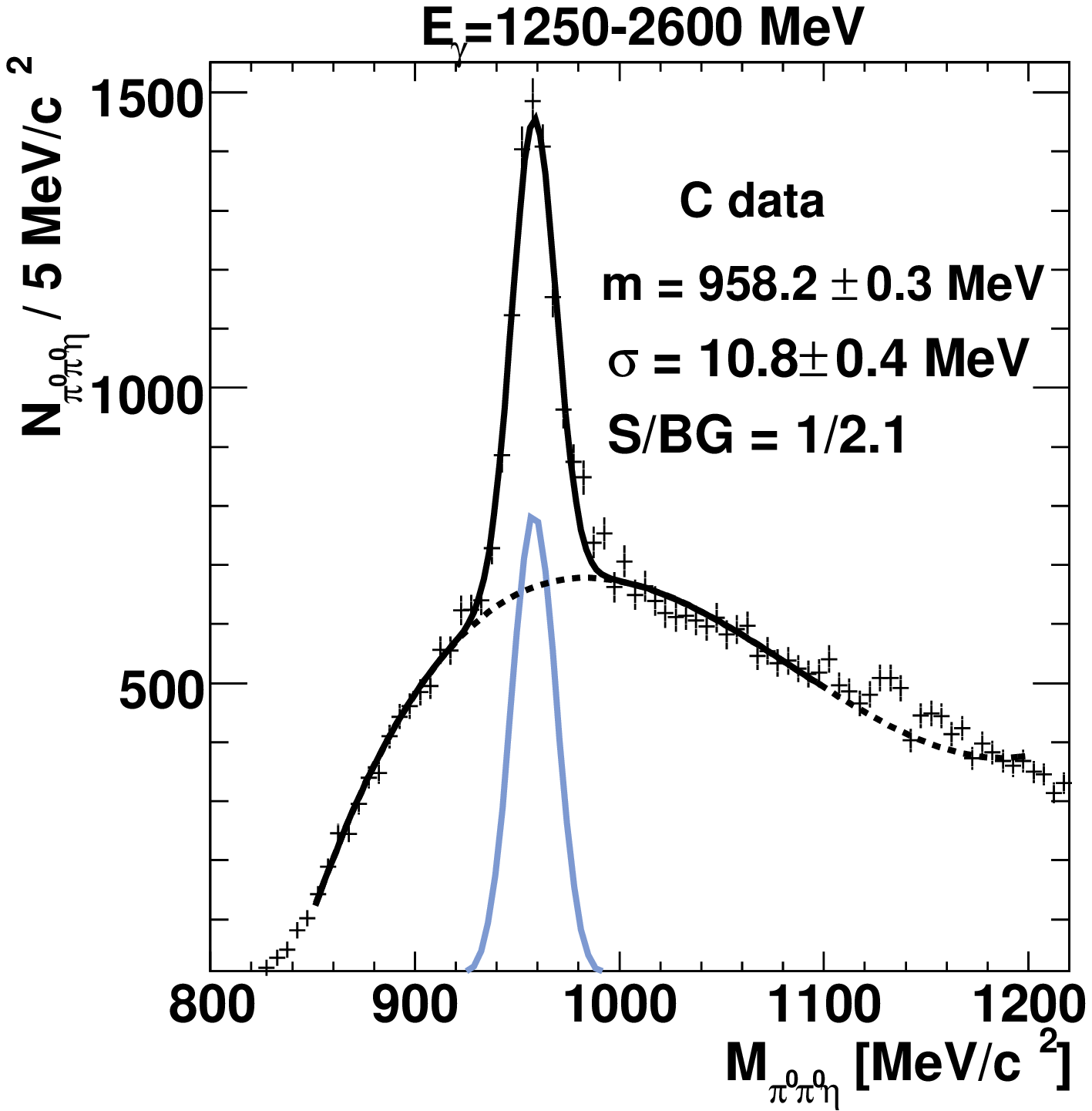}
  }
\caption{(Color online)  Left:  Setup used in the experiment in 2009. Right: The $\pi^0\pi^0\eta$ invariant mass distribution measured in photoproduction off carbon in the incident photon energy range of 1250-2600 MeV. The solid curve represents a fit to the data using a Gaussian function combined with a polynomial function for the background. The fit parameters are: $\sigma$=10.8$\pm$0.4 MeV (corresponding to the instrumental resolution), m=958.2$\pm$0.3 MeV/$c^2$; S/BG is the signal (S) to background (BG) ratio within a $\pm$3$\sigma$ interval.}
\label{fig:setup}
\end{figure}
\par
The Crystal Barrel detector, a homogeneous electromagnetic calorimeter, consisted of 1230 Cs(Tl) crystals read out with photodiodes, and was arranged in 21 rings, subtending polar angles of 30$^\circ$ - 156$^\circ$. A three layer  fibre detector (Inner Detector) with 513 scintillating fibres for charged particle detection~\cite{Suft} surrounded the target, placed at the centre of the CB. In the forward angular range between 12$^\circ$ and 28$^\circ$, 90 CsI(Tl) crystals (Forward Plug FP) were mounted and read out with photomultipliers (PMT) providing energy and time information and equipped with plastic scintillators for charged particle detection. The angular range between 1$^\circ$ and 12$^\circ$  was subtended by the MiniTAPS forward wall at a distance of 235 cm from the centre of the CB, comprising 216 BaF$_2$ crystals read out via PMTs with a readout electronics described in~\cite{Drexler}. Each BaF$_2$ module was equipped with a plastic scintillator for charged particle identification. To suppress electromagnetic background at forward angles and for a better separation between charged pions and photons an aerogel Cherenkov detector (n=1.05) was placed in front of MiniTAPS. A schematic view of the main detectors is shown in Fig.~\ref{fig:setup} (left). The high granularity of this system makes it very well suited for the detection of multi-photon final states. Downstream of the MiniTAPS detector, a Gamma Intensity Monitor (GIM) was located for measuring the photon flux needed for the absolute normalisation of the cross sections~\cite{Jan}. A detailed description of the full detector setup is given in~\cite{Karoly}.\\
\par 
A solid target of natural carbon, with a 30 mm diameter and thickness of 15 mm, corresponding to $\approx$8\% of the radiation length $X_{0}$, was mounted in the centre of the CB detector. For a reference measurement, data taken in November 2008 on a liquid hydrogen target (50 mm length and 30 mm diameter) were analysed. \\
\par
Online event selection was made using first- and second-level triggers. The detectors contributing to the first-level trigger were the FP and MiniTAPS together with signals from the tagger. Due to the long rise time of the photodiode signals, CB could not be used in the first-level trigger. The second-level trigger was based on a FAst Cluster Encoder (FACE), providing the number of clusters in the CB within $\approx $ 10 $\mu$s. Overall a cluster multiplicity of 4 was required in MiniTAPS, FP and CB~\cite{Stefan}.  \\
\begin{figure}
 \resizebox{1.\textwidth}{!}
  {
   \includegraphics[height=1.\textheight]{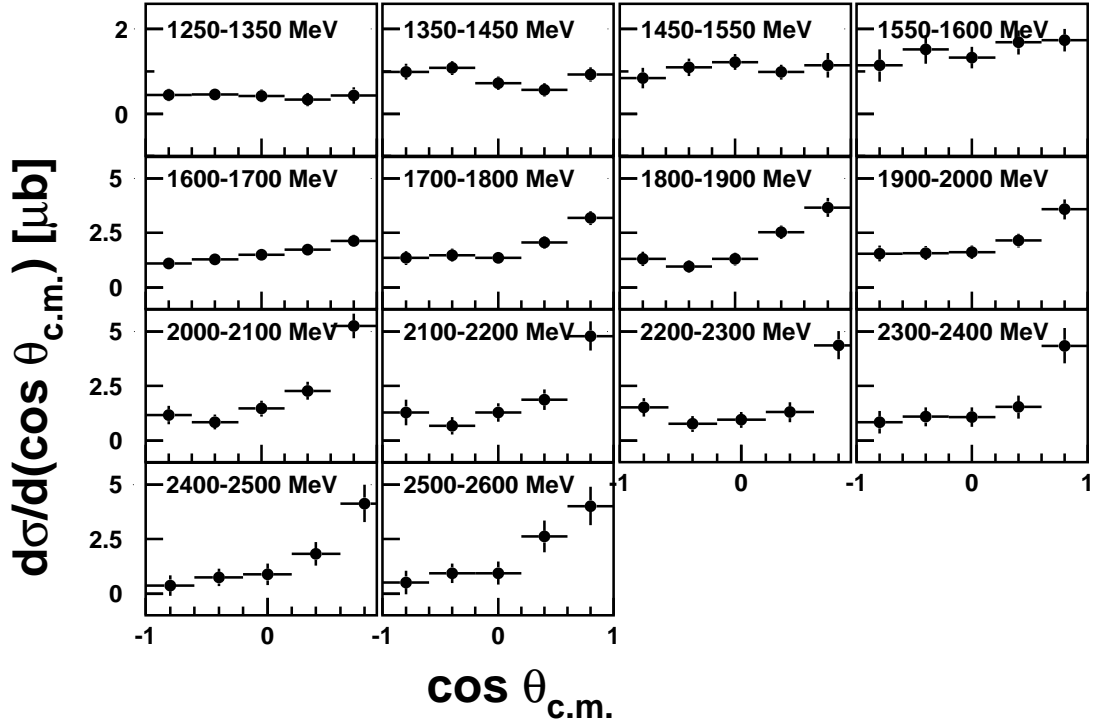} 
     }
\caption{Differential cross sections for photoproduction of $\eta^\prime$ mesons off C for different bins in the incident photon energy range 1250-2600 MeV determined in 5 $\cos\theta^{\eta'}_{c.m.}$ bins of width 0.4.}
\label{fig:diffcs}
\end{figure}
\section{Data Analysis}
\label{sec:ana}
The $\eta^\prime$ mesons were identified in the neutral decay channel $\eta^\prime\rightarrow \pi^{0}\pi^{0}\eta \rightarrow 6 \gamma$ with a branching ratio of 8.1\%. For the reconstruction, only events were selected with 6 neutral and 0 or 1 charged hits and with an energy sum of neutral clusters higher than 600 MeV. The 6 photons were combined in 2 pairs of 2 photons with invariant masses in the range 110 MeV/c$^2 \le m_{\gamma\gamma} \le$ 160 MeV/c$^2$ (close to $m_{\pi^{0}}$) and one pair with invariant mass in the range 500 MeV/c$^2 \le m_{\gamma\gamma} \le$ 600 MeV/c$^2$ (close to $m_{\eta}$). The 6$\gamma$ events with 3 pairs with invariant masses close to the pion mass ($m_{\pi^{0}}$) were rejected from the data set in order to suppress the background coming from $\eta \rightarrow 3\pi^{0}$ decays. The resulting $\pi^{0}\pi^{0}\eta$ invariant mass spectrum is shown in Fig.~\ref{fig:setup} (right). In total, 4300 $\eta^\prime$ mesons were reconstructed in the photon energy range 1250 - 2600 MeV. 80 $\eta^\prime$ mesons were identified in events with 7 neutral hits, corresponding to $\eta^\prime$ photoproduction on a neutron with the recoil neutron registered in the detector. Because of the very poor signal-to-background ratio (1/25), these events were not included in the further analysis. The spectrum was fitted with a Gaussian and a background function: $f(m) = a \cdot (m-m_1)^{b} \cdot (m-m_2)^{c} $ or a 3rd order polynomial. As it can be seen from the fit parameters in Fig.~\ref{fig:setup} (right) the $\eta^\prime$-signal in the $\pi^{0}\pi^{0}\eta$ spectrum has a width $\sigma$=$10.8\pm0.4$ MeV and a position $m=958.2\pm0.3$ MeV/$c^2$. The same fit procedure was used to determine the $\eta^\prime$ yield in 14 bins of the incident photon energy and in 5 bins of $\cos\theta^{\eta^\prime}_{c.m.}$, where $\theta^{\eta^\prime}_{c.m.}$ is the angle of the $\eta^\prime$ in the centre of mass system of the incident photon and a target nucleon {\it at rest}. The bins were chosen considering the statistics available in this measurement. The total cross section for $\eta^\prime$ photoproduction was extracted via two independent methods, namely by integrating the differential cross sections and by direct reconstruction of the $\eta^\prime$ meson in bins of 50 MeV in the incident photon energy  range 1250-2600 MeV. The results from both methods are compared and further discussed in Sec. \ref{sec:tot}.\\
\par 
For the cross section measurement the acceptance for reconstructing the reaction of interest was determined. Monte Carlo (MC) simulations of the reaction $\gamma C
\rightarrow X  \eta^\prime$ were performed for the solid carbon target with the GEANT3 package, using as input the angular distribution of the $\eta^\prime$ meson deduced from the experimental data and taking the Fermi motion of nucleons in the target nucleus into account. The reconstruction of simulated $\eta^ \prime$ data was performed for the
same trigger conditions as in the experiment and for the same incident photon energy range from 1250 to 2600 MeV.\\
The statistical errors were determined from the yield of the $\eta^\prime$ signal in each energy and $\cos(\theta^{\eta^\prime}_{c.m.})$ bin (S) and the counts in the background under the peak (BG) according to the formula: $\Delta N = \sqrt{(S+2BG)}$.\\
\begin{figure}
 \resizebox{1.\textwidth}{!}
  {
   \includegraphics[height=0.8\textheight]{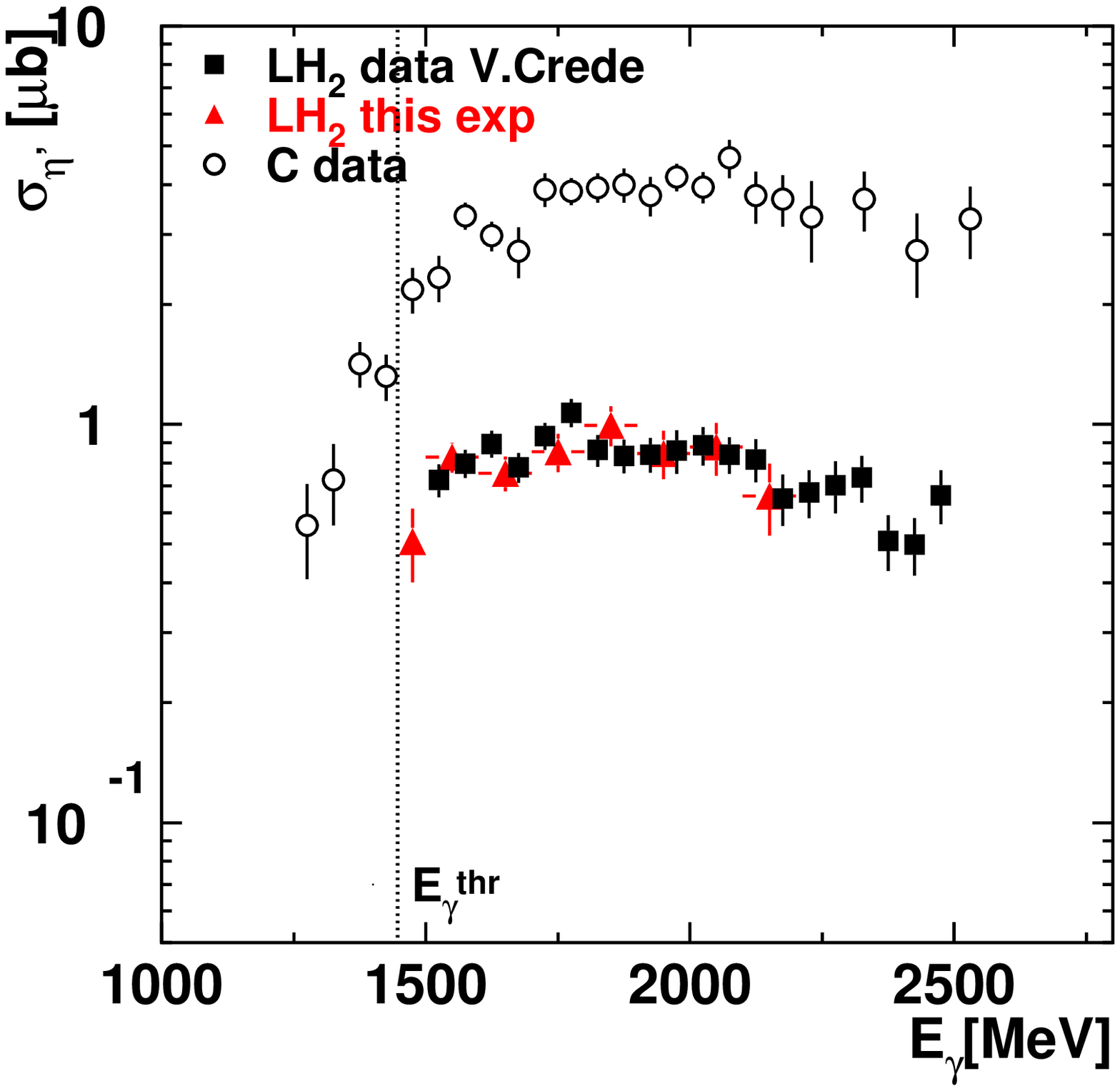} \includegraphics[height=0.8\textheight]{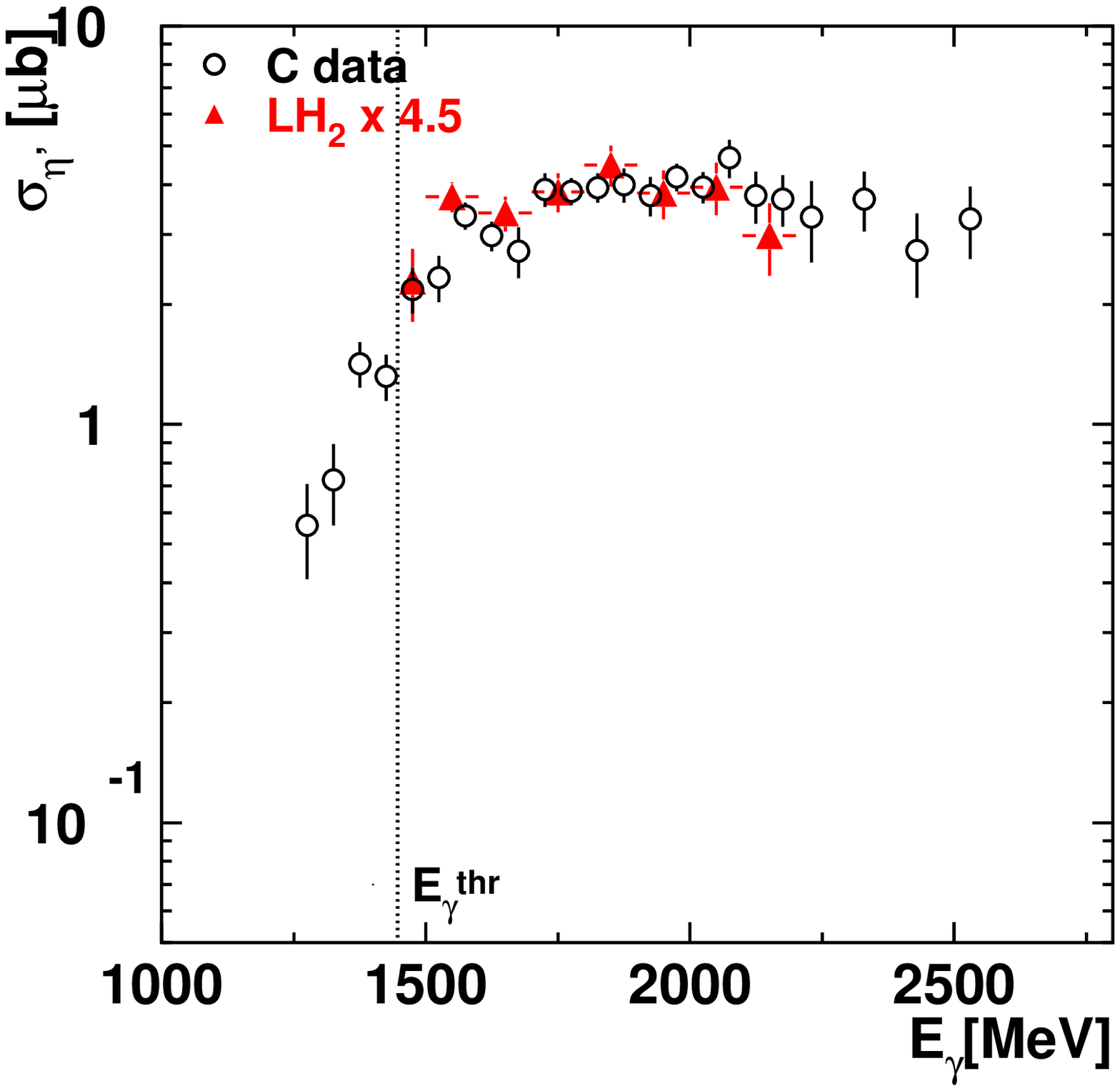} 
     }
\caption{(Color online) Left: Total cross sections for $\eta^\prime$ photoproduction off C (open points), in comparison to those for $\eta^\prime$ production off the proton from this experiment (triangles) and from \cite{Crede} (full squares). Right: Total cross section for $\eta^\prime$ production off the proton from this experiment scaled up by factor of 4.5 to match the cross section for $\eta^\prime$ production off C.}
\label{fig:tot_p}
\end{figure}
\par 
The different sources of systematic errors are summarized in Table \ref{tab:syst}. Applying different background functions, the systematic errors in the fit procedure were estimated to be in the range of 10-15\%. Systematic errors in the acceptance determination were investigated by varying the start distributions in the acceptance simulation between isotropic and forward peaking $\eta^\prime$ angular distributions, and were found to be less than 10\%. The photon flux through the target was determined by counting the photons reaching the GIM in coincidence with electrons registered in the tagger system. The systematic errors introduced by the photon flux determination were estimated to be about 5-10 $\%$. Systematic errors of $\approx 5\%$ came from uncertainties in the effective number of participating nucleons seen by the incident photons due to photon shadowing ~\cite{bianchi}. Adding the systematic errors quadratically, the total systematic error in the determination of  the cross sections was of the order of 20$\%$.
\begin{table}[h!]
\centering
\caption{Sources of systematic errors}
\begin{footnotesize}

\begin{tabular}{cc}
\hline
fits & $\approx 10-15\%$\\
acceptance & $\lesssim 10\%$\\
photon flux & 5-10 $\%$\\
photon shadowing & $\approx 5\%$\\
total& $\approx 20\%$\\
\hline
\end{tabular}
\end{footnotesize}

\label{tab:syst}
\end{table}

\section{Experimental results}
\subsection{Differential cross sections for the $\eta^\prime$ photoproduction off carbon}
\label{sec:diff}
The differential cross sections have been determined according to:
\begin{equation}
\frac{d\sigma}{d (\cos \theta^{\eta^\prime}_{c.m.})} = \frac{N_{\eta^\prime\rightarrow \pi^0 \pi^0 \eta}}{A_{\eta^\prime\rightarrow \pi^0 \pi^0 \eta}} \cdot \frac{1}{N_{\gamma} \cdot n_{t}} \cdot \frac{1}{\Delta \cos\theta^{\eta^\prime}_{c.m.}} \cdot \frac{1}{\frac{\Gamma_{\eta^\prime \rightarrow \pi^0 \pi^0 \eta \rightarrow 6 \gamma}}{\Gamma_{total}}},
\end{equation}
where $N_{\eta^\prime \rightarrow \pi^0 \pi^0 \eta}$ is the number of reconstructed $\eta^\prime$ mesons extracted by the fit procedure as described in Sec.~\ref{sec:ana} in each ($E_{\gamma}$, $\cos\theta^{\eta^\prime}_{c.m.}$) bin; $A_{\eta^\prime \rightarrow \pi^0 \pi^0 \eta}$ is the acceptance in each  ($E_{\gamma}$, $\cos\theta^{\eta^\prime}_{c.m.}$) bin; $N_{\gamma}$ is the number of photons in an E$_\gamma$ bin; $n_{t}$ is the density of the target nucleons multiplied by the target thickness (2.15 $\cdot$10$^{23}$ cm$^{-2}$ (for the LH$_{2}$ target) and 1.26 $\cdot$10$^{23}$ cm$^{-2}$ (for the C target), respectively); $\Delta \cos\theta^{\eta^\prime}_{c.m.}$ is the angular bin in the c.m. frame; $\frac{\Gamma_{\eta^\prime \rightarrow \pi^0 \pi^0 \eta \rightarrow 6 \gamma}}{\Gamma_{total}}$ is the decay branching fraction of 8.1\% for the decay channel $\eta^\prime \rightarrow \pi^0 \pi^0 \eta \rightarrow 6 \gamma$.\\
\par 
The differential cross sections $d\sigma /d (\cos\theta^{\eta^\prime}_{c.m.})$ are presented in Fig.~\ref{fig:diffcs} for 14 bins in the incident photon energy range. A rather flat angular distribution is observed at low energies - below and at the production threshold on a free nucleon ($E_{\gamma}^{thr}$=1447 MeV). For higher photon energies $E_{\gamma} \ >$  1700 MeV, the angular distribution shows a peaking in the forward direction, characteristic for t-channel production. This behaviour is similar to previous results on angular distributions for $\eta^\prime$ photoproduction off the proton and off the deuteron~\cite{Crede,Igal,Dugger,Williams}.

\subsection{Total cross section for the $\eta^\prime$ photoproduction off carbon}
\label{sec:tot}
In Fig.\ref{fig:tot_p} the total cross section for the $\eta^\prime$ photoproduction off carbon is compared to the total cross section for $\eta^\prime$ meson production off the proton measured in this experiment and in the previous measurement \cite{Crede}, respectively.
The comparison shows that the experimental data for $\eta^\prime$ production off the proton from the present measurement are in good agreement with the measured total cross sections reported in \cite{Crede} (Fig.\ref{fig:tot_p} left), providing an independent check of the data analysis. While the $\eta^\prime$ cross section drops dramatically near the production threshold of $E_{\gamma}$=1447 MeV in case of the proton target (Fig.~\ref{fig:tot_p} left), there is appreciable yield below this threshold in the reaction on the C target. Above threshold, the shapes of the excitation functions for both targets are similar, as shown in Fig.~\ref{fig:tot_p} right where the LH$_{2}$ data are scaled up by a factor of 4.5. \\
\par 
There are several effects which can cause a non-zero cross section below 1447 MeV for the C target. On the one hand, the nucleons in the target are not at rest but have some Fermi momentum which gives rise to a distribution of the energy $\sqrt s$ available in the centre-of-mass system for a given incident photon energy. On the other hand, due to inelastic reactions in the target nucleus, the lifetime of the $\eta^\prime$ meson may be reduced and its width correspondingly increased, as  measured in~\cite{nanova}. The resulting tail in the mass distribution allows a production below the free threshold energy. In addition, as discussed in the next section, also the mass of the meson might drop in a nuclear medium which lowers the production threshold and again makes subthreshold production possible.
\begin{figure}
 \resizebox{1.\textwidth}{!}
  {
   \includegraphics[height=0.9\textheight]{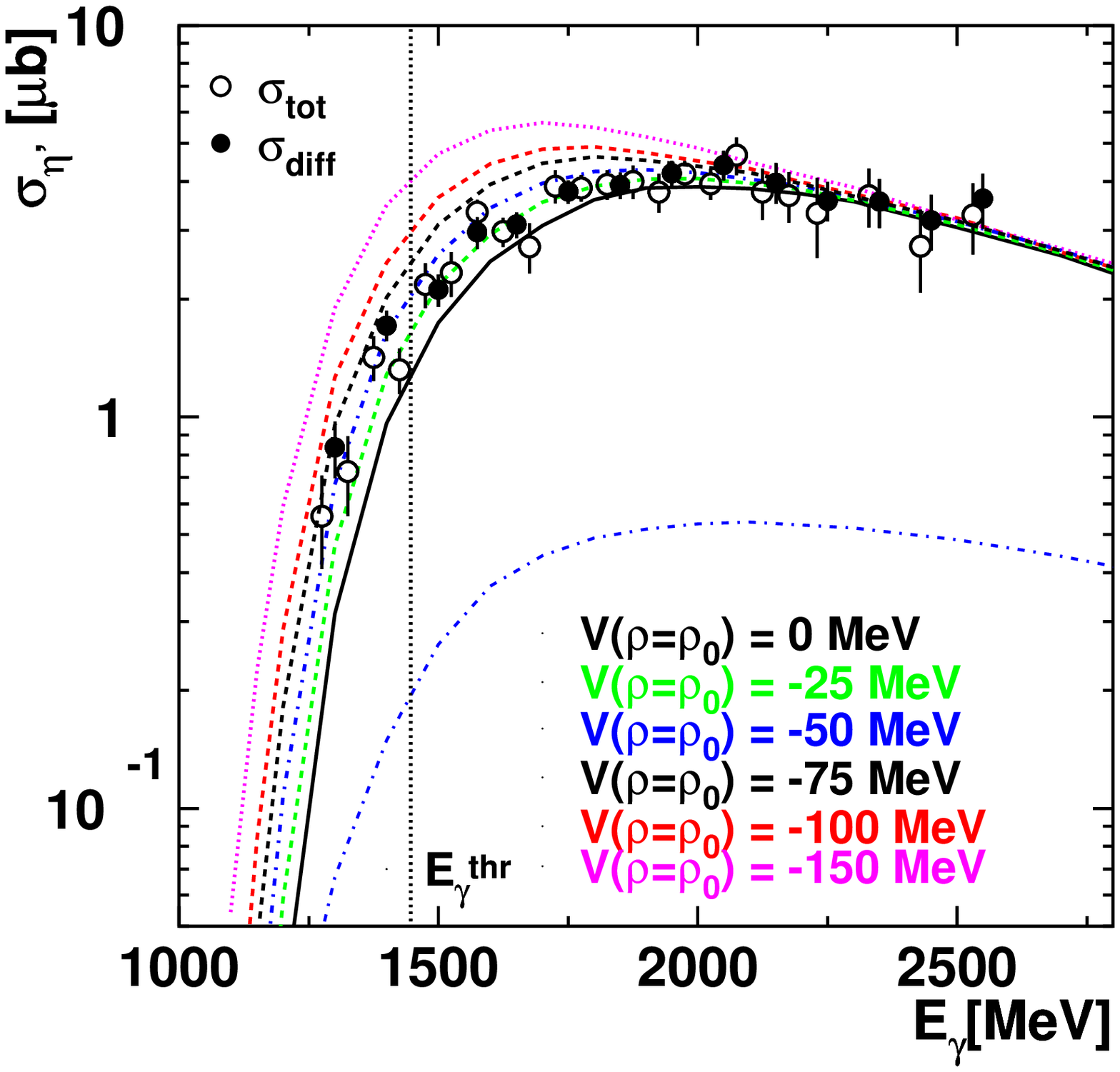}  \includegraphics[height=0.9\textheight]{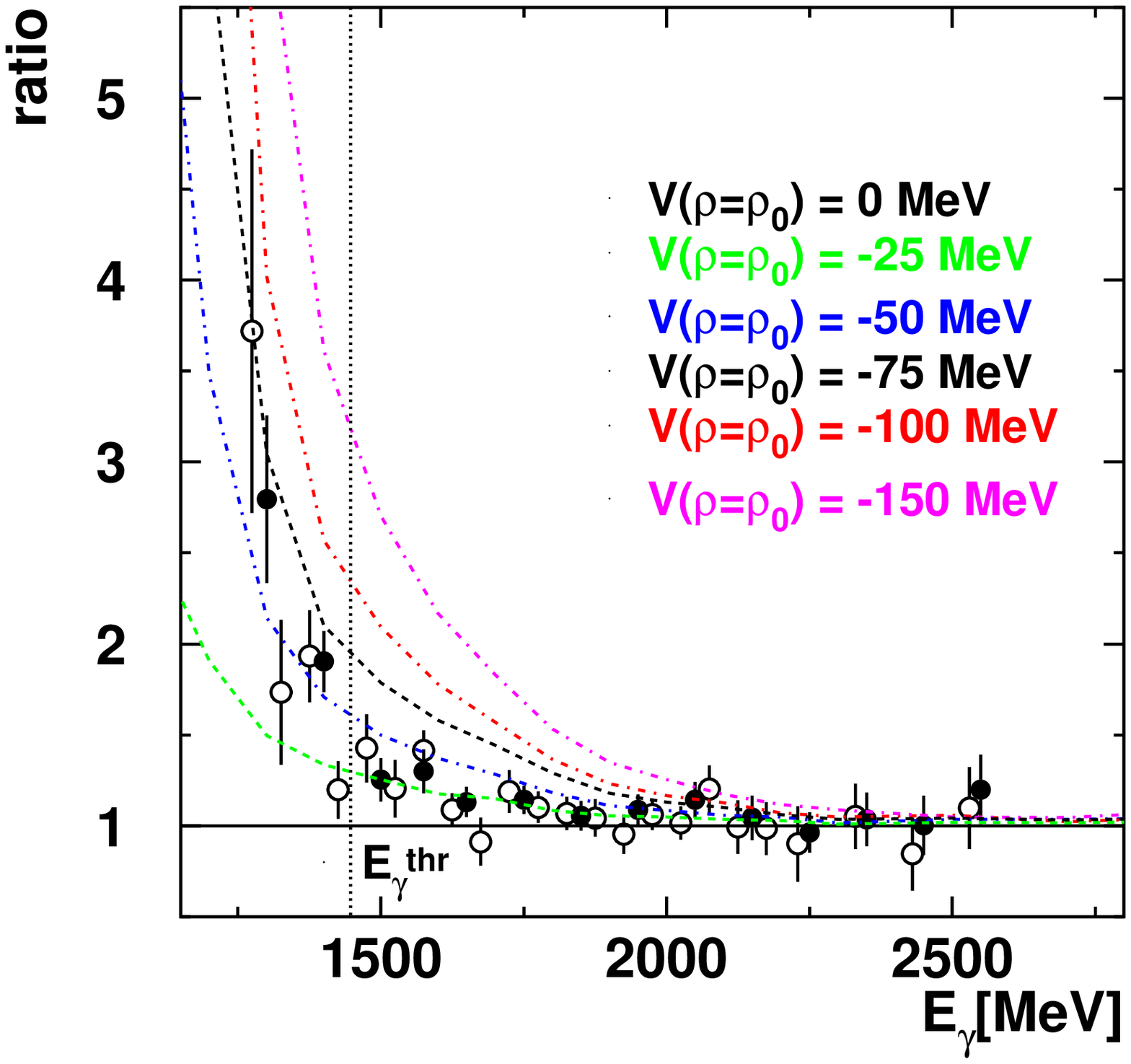} \includegraphics[height=0.9\textheight]{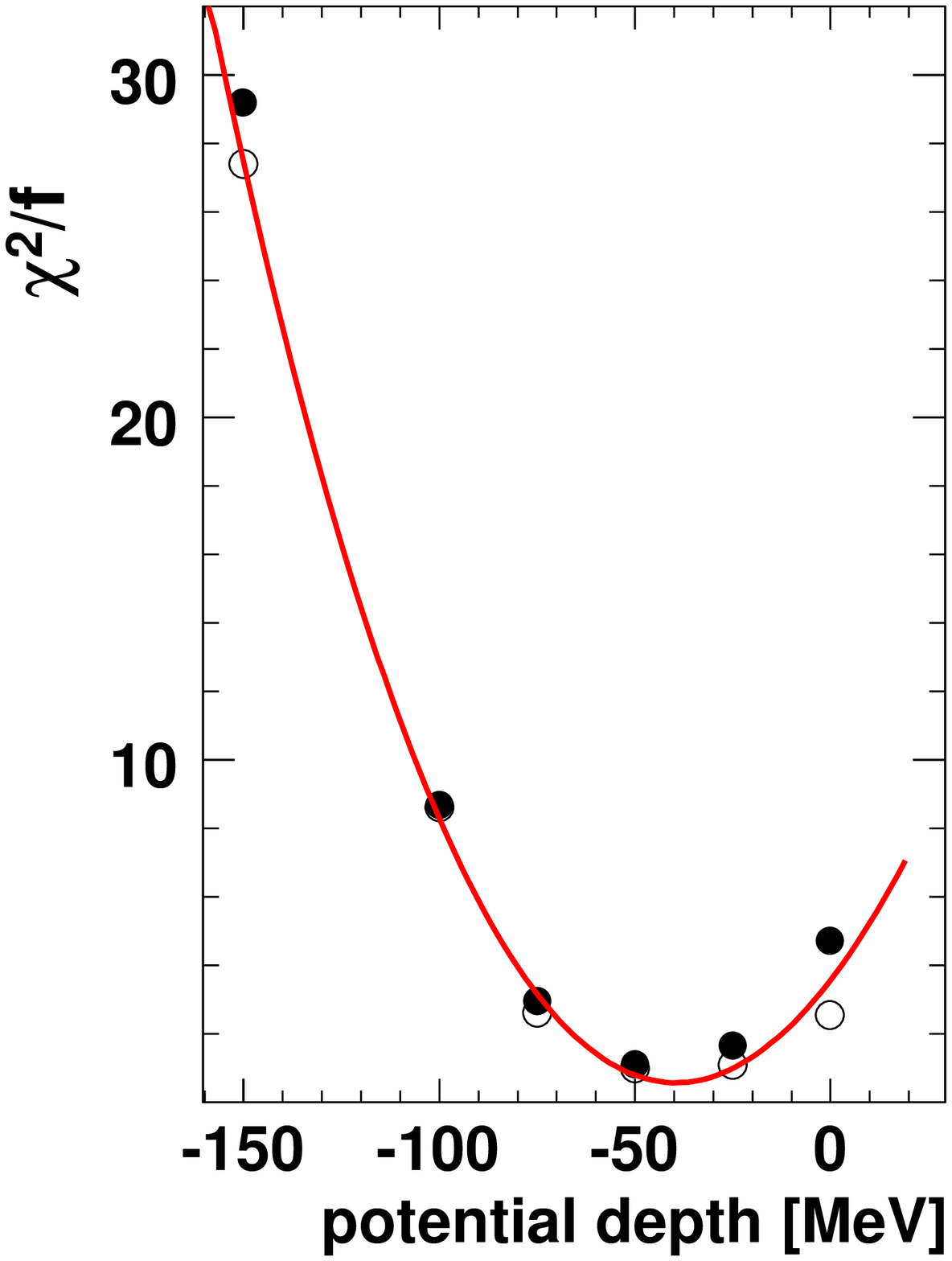} 
     }
\caption{(Color online) Left: Total cross section for $\eta^\prime$ photoproduction off C. The experimental data are extracted by integrating the differential cross sections (full circles) and by direct measurement of the $\eta^\prime$ yield in the incident photon energy bins of width $\Delta E_{\gamma}$=50 MeV (open circles). The calculations are for $\sigma_{\eta^\prime N}$=11 mb and for potential depths $V$=0 MeV (black line), -25 MeV (green), -50 MeV (blue), -75 MeV (black dashed), -100 MeV (red) and -150 MeV (magenta) at normal nuclear density, respectively, and using the full nucleon spectral function. The dot-dashed blue curve is calculated for correlated intranuclear nucleons only (high-momentum nucleon contribution). All calculated cross sections have been reduced by a factor 0.75 (see text). Middle: The experimental data and the predicted curves for $V$=-25, -50, -75, -100 and -150 MeV divided by the calculation for scenario of $V$=0 MeV and presented on a linear scale. Right: $\chi^{2}$-fit of the data with the calculated excitation functions for the different scenarios over the full incident photon energy range. }
\label{fig:exc}
\end{figure}

\begin{figure}
 \resizebox{1.\textwidth}{!}
  {
   \includegraphics[height=0.8\textheight]{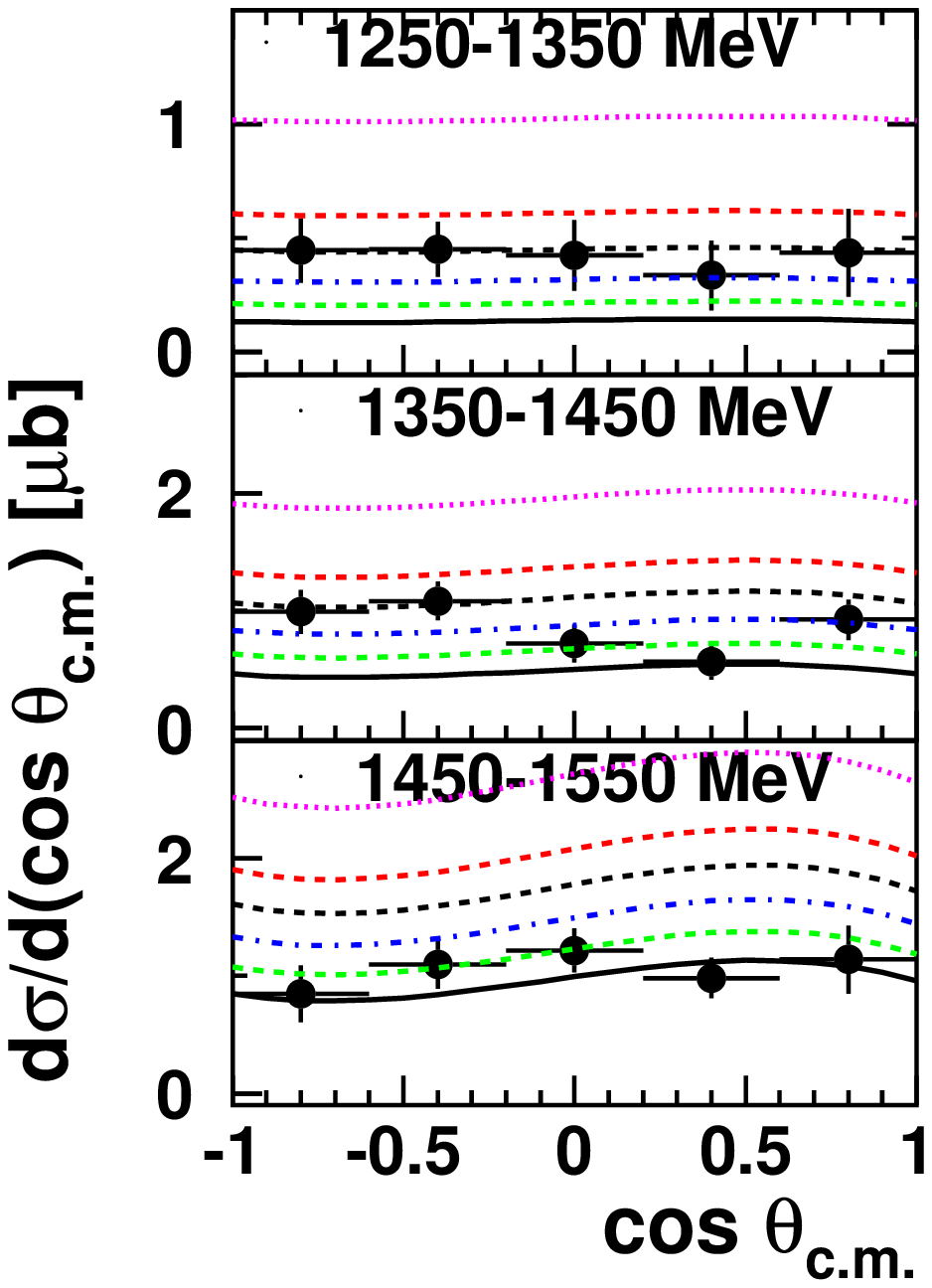} \includegraphics[height=0.8\textheight]{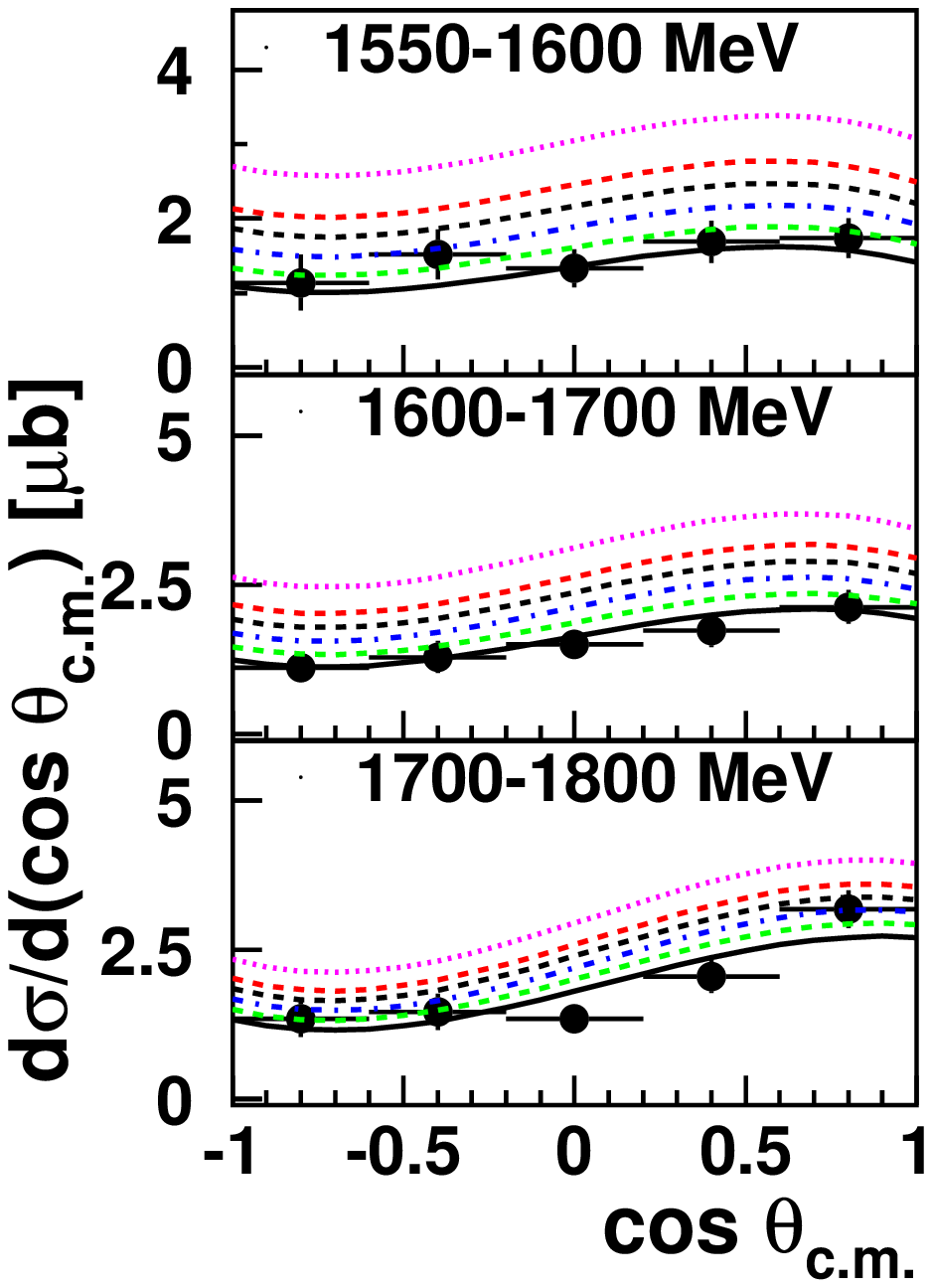}\includegraphics[height=0.8\textheight]{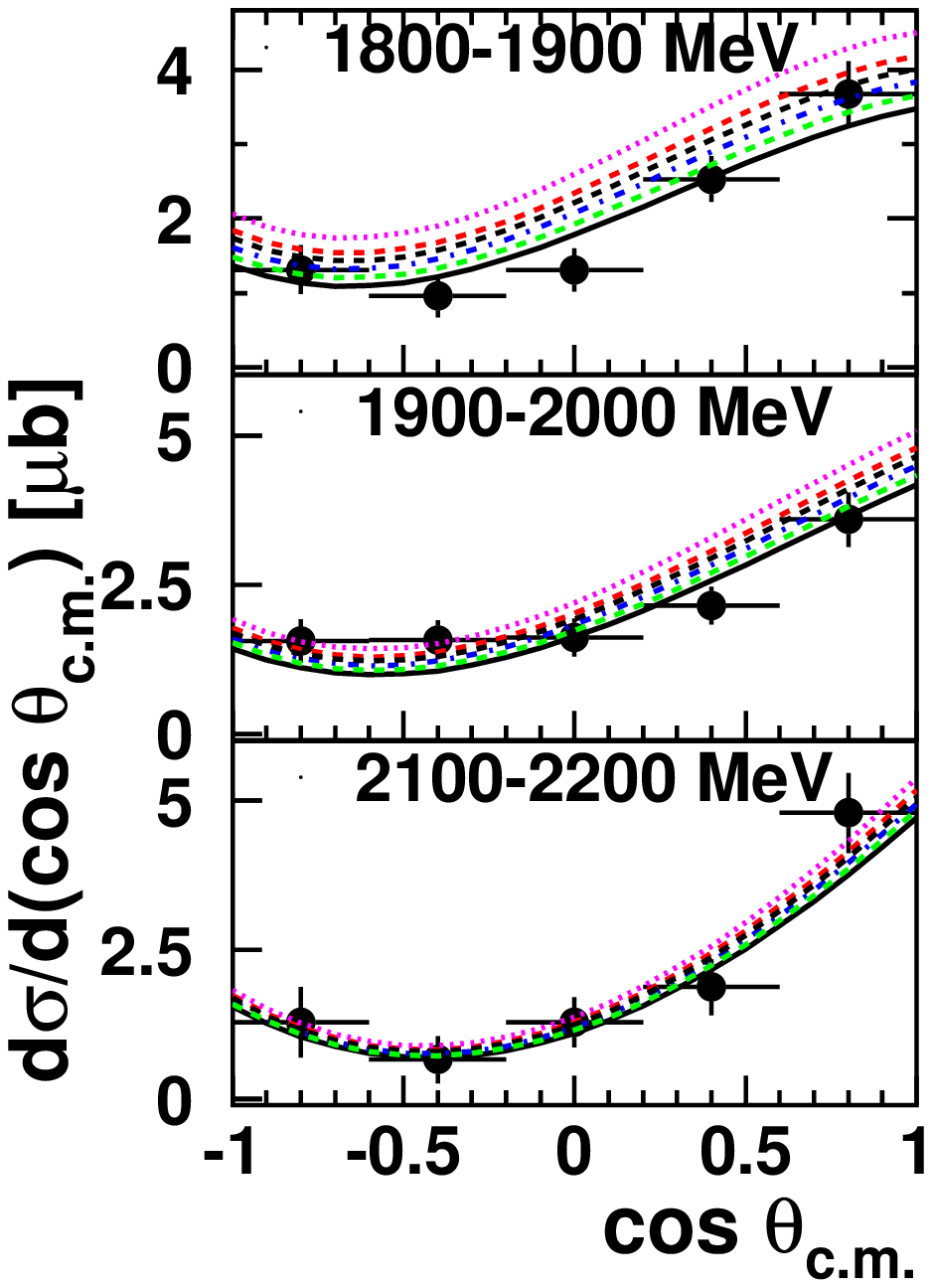} 
     }
\caption{(Color online) Differential cross sections for $\eta^\prime$ photoproduction off C for incident photon energies below the free production threshold (left), at the threshold (middle), and above the threshold (right). The calculations are for $\sigma_{\eta^\prime N}$=11 mb and for potential depths $V$=0, -25, -50, -75, -100 and -150 MeV, at normal nuclear density, respectively. All calculated cross sections have been reduced by a factor 0.75 (see text). The colour code is identical to the one in Fig.\ref{fig:exc}.}
\label{fig:diff_c}
\end{figure}
\section{Comparison to the theoretical model predictions}
\subsection{Excitation function for  $\eta^\prime$ mesons}
\label{sec:excit}
\par
In Fig.~\ref{fig:exc} (left) the measured excitation function for photoproduction of $\eta^\prime$ mesons off carbon
       is compared to calculations within the first collision model based on the nucleon spectral
      function and described in detail in \cite{Paryev}. Starting from the measured differential cross sections
     for $\eta^\prime$ production off  the ÒfreeÓ proton and neutron \cite{Crede,Igal}, the cross section for $\eta^\prime$
     photoproduction off carbon is calculated in an eikonal approximation, taking  the effect of the
     nuclear $\eta^\prime$ mean-field potential into account. Here,  the off-shell differential cross sections on
     the intranuclear proton and neutron for the production of an $\eta^\prime$ meson with reduced in-medium
     mass in the elementary reactions $\gamma p \rightarrow \eta^\prime p$ and $\gamma n \rightarrow \eta^\prime n$ are assumed to be given by the measured
     on-shell cross sections, applying off-shell kinematics. The $\eta^\prime$ final-state absorption is determined by
     the inelastic in-medium $\eta^\prime N$ cross section taken to be $\sigma_{inel}$=11 mb, consistent with the result of
     transparency ratio measurements \cite{nanova}.  The total nucleon spectral function is used in the 
     parametrization given in \cite{Efremov}. Thereby, the contribution of $\eta^\prime$ production from two-nucleon 
     short-range correlations is taken into account. The calculations are improved with respect to \cite{Paryev} as
     the momentum-dependent optical potential from \cite{Rudy}, seen by the nucleons emerging from the nucleus in coincidence with the $\eta^\prime$ mesons, is 
     accounted for as well. 
\par
The calculations have been performed for six different scenarios assuming depths of the $\eta^\prime$ real potential at normal nuclear matter density of $V$=0, -25, -50, -75, -100 and -150 MeV, respectively. To correct for the absorption of incident photons, not considered in the calculations, the predicted cross sections have been scaled down by 10\% according to~\cite{bianchi}. The calculated cross sections have been further scaled down - within the limits of the systematic uncertainties - by a factor of 1.2 to match the experimental excitation function data at incident photon energies above 2.2 GeV, where the difference between the various scenarios is very small. In Fig.~\ref{fig:exc} (middle) the experimental data and the calculations for the different scenarios are divided by the calculation for $V$=0 MeV and are presented on a linear scale. The data follow the general upward trend of the calculated cross section ratios towards lower incident energies. The highest sensitivity to the $\eta^\prime$ potential depth is given for incident photon energies below the production threshold on the free nucleon, however, there, the statistical errors become quite large. It is nevertheless seen from Fig.~\ref{fig:exc} left and Fig.~\ref{fig:exc} middle that the excitation function data appear to be incompatible with $\eta^\prime$ mass shifts of -100 MeV and more at normal nuclear matter density. A $\chi^{2}$-fit of the data with the calculated excitation functions for the different scenarios (see Fig.~\ref{fig:exc} right) over the full range of incident energies gives a potential depth of -(40$\pm$6) MeV.\\
\par
It has been investigated whether the observed cross section enhancement relative to the $V$=0 MeV case could also be due to $\eta^\prime$ production on dynamically formed compact nucleonic configurations - in particular, on pairs of correlated nucleon clusters - which share energy and momentum. These effects have been studied experimentally \cite{Debowski} and theoretically \cite{Sibirtsev, Paryev_kaon} in very near-threshold $K^+$ production in proton-nucleus reactions and can be taken into account - as has been done in the present calculations - by using the full nucleon spectral function including high momentum tails. Applying the parametrization of the spectral function given by~\cite{Efremov}, Fig.~\ref{fig:exc} left shows that correlated high momentum nucleons contribute only about 10-15\% to the $\eta^\prime$ yield in the incident energy regime above 1250 MeV. The observed cross section enhancement can therefore be attributed mainly to the lowering of the $\eta^\prime$ mass in the nuclear medium.\\
\par 
A real part of the $\eta^\prime$-nucleus potential depth between -75 and -25 MeV is confirmed by comparing the experimental angular distributions with the corresponding calculations. Fig.~\ref{fig:diff_c} shows a comparison for incident photon energy ranges  below, at and above the free production threshold, respectively. As for the excitation function, the highest sensitivity to the potential depth is found for low incident energies, while at higher energies the measured angular distributions are reproduced quite well by the calculations independent of the assumed potential depth. 
\begin{figure}
 \resizebox{1.\textwidth}{!}
  {
   \includegraphics[height=0.8\textheight]{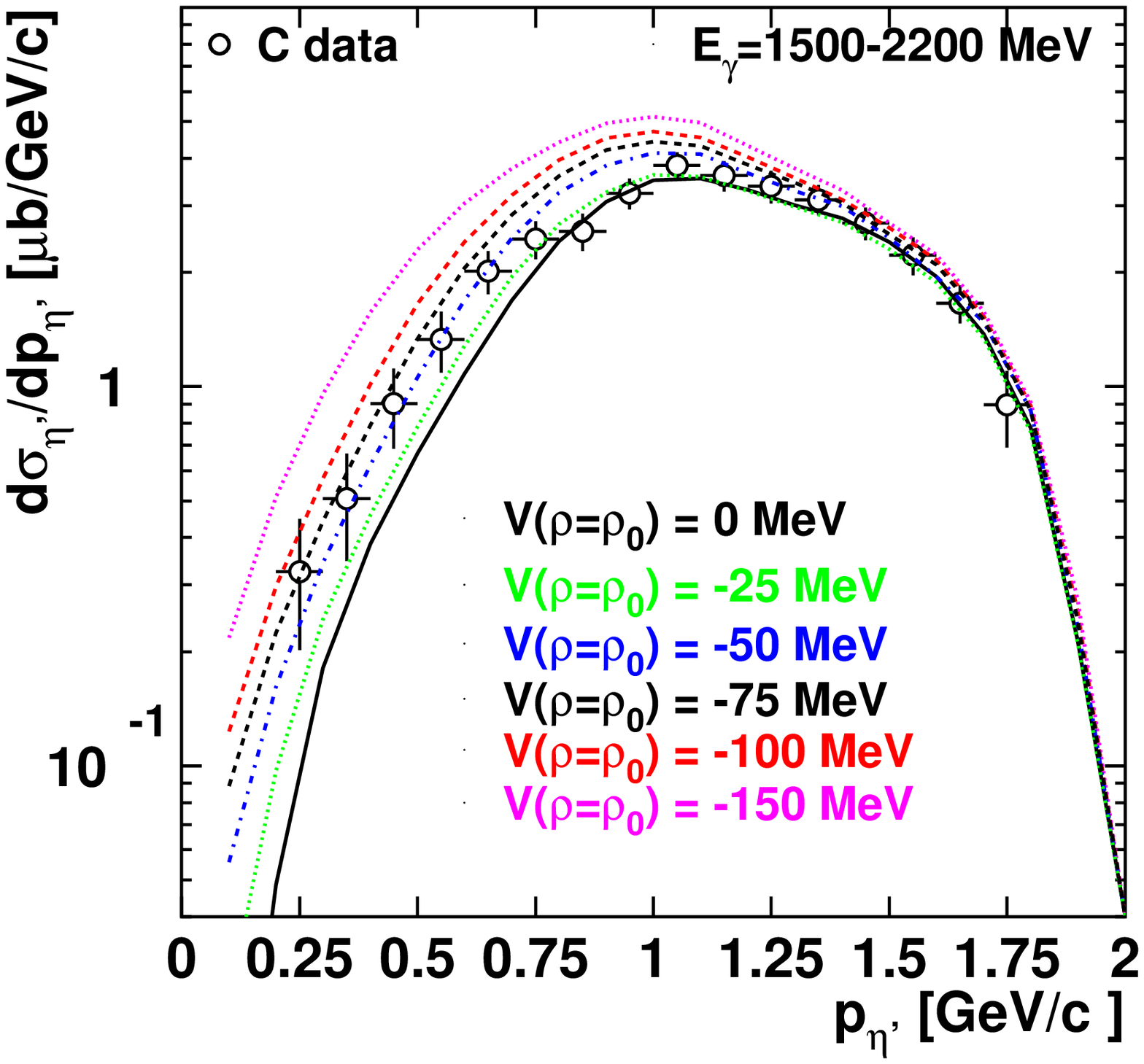}  \includegraphics[height=0.8\textheight]{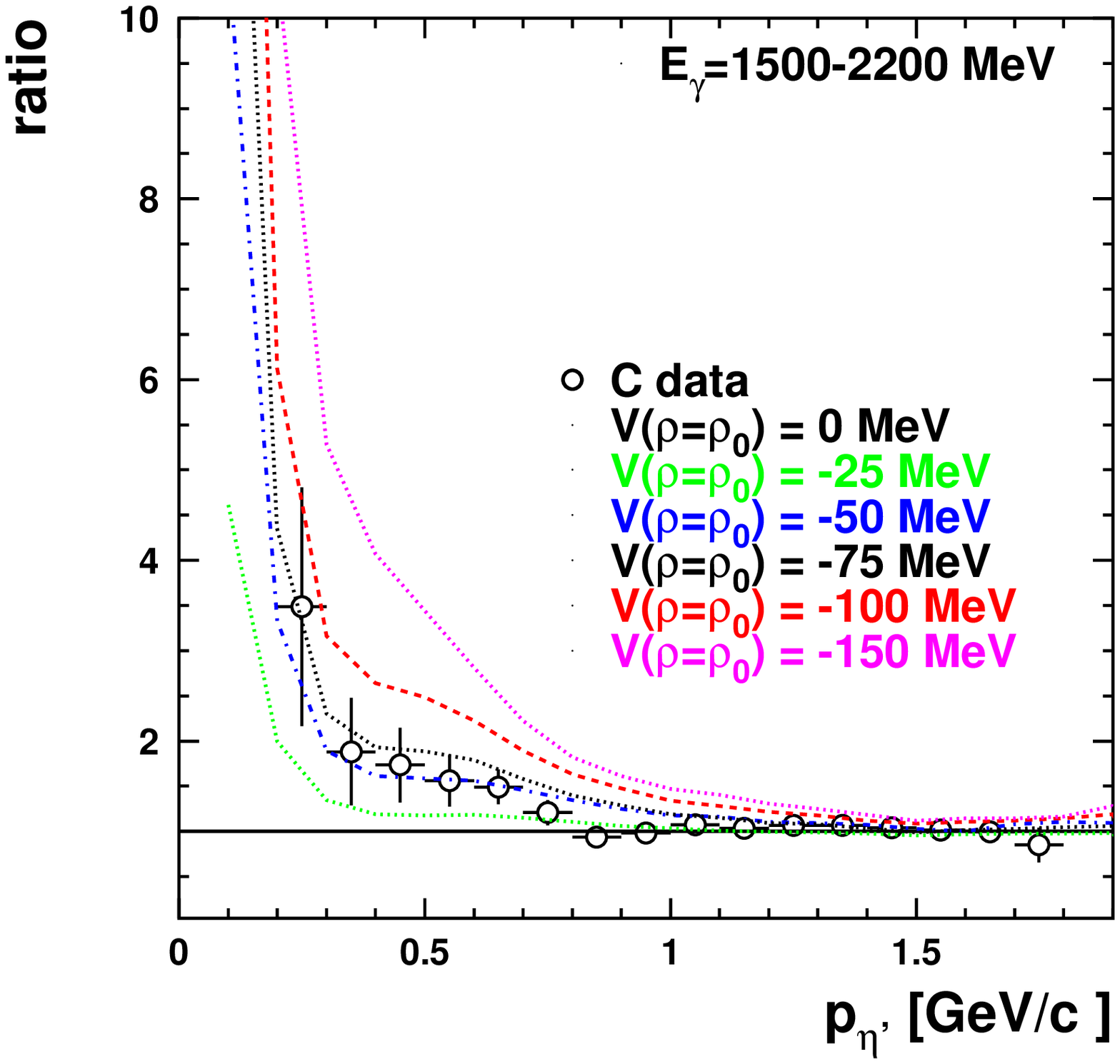} \includegraphics[height=0.8\textheight]{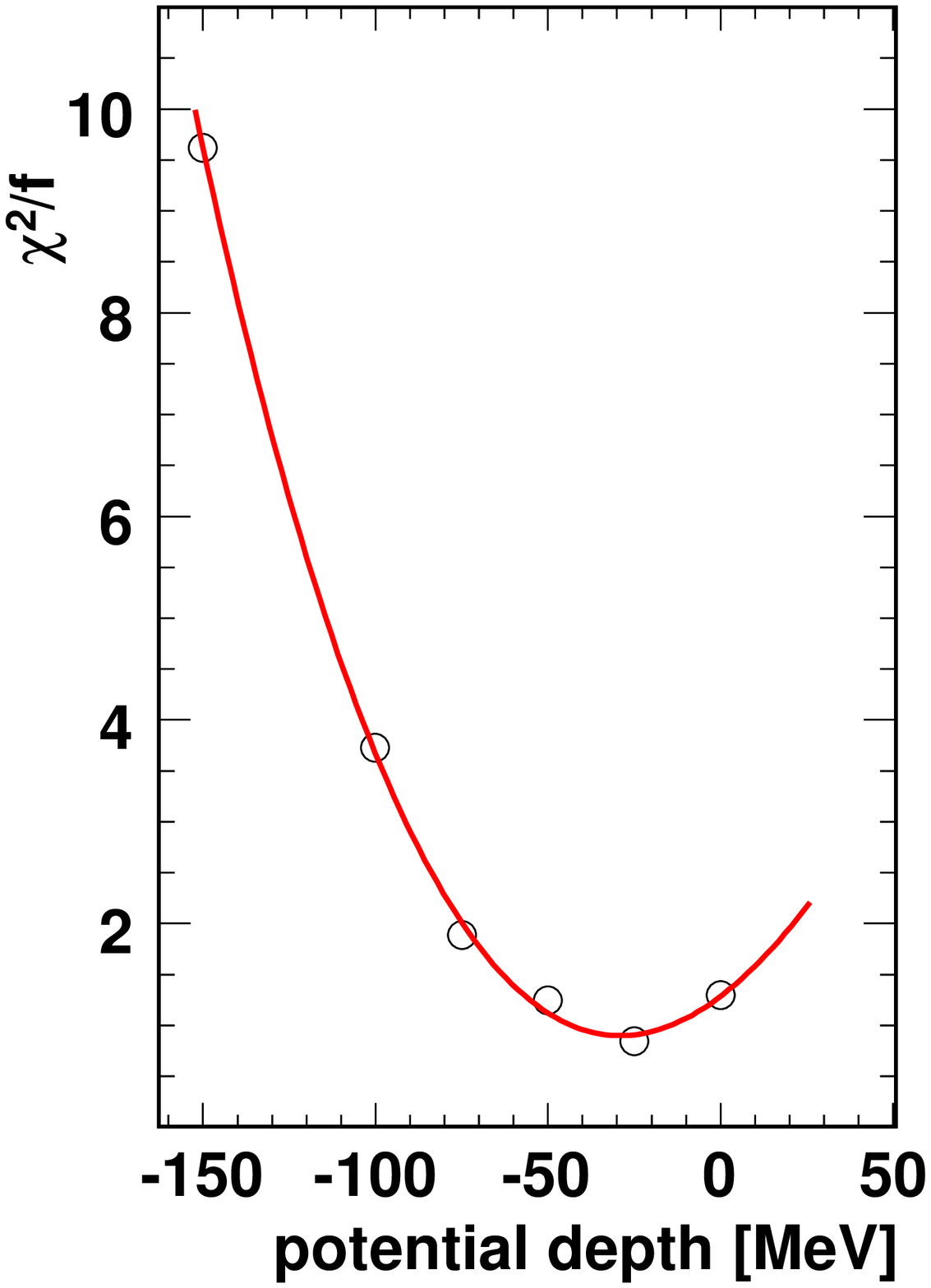} 
     }
\caption{(Color online) Left: Momentum distribution for $\eta^\prime$ photoproduction off C for the incident photon energy range 1500-2200 MeV. The calculations are for  $\sigma_{\eta^\prime N}$=11 mb and for potential depths $V$=0, -25, -50, -75, -100 and -150, at normal nuclear density, respectively. All calculated cross sections have been reduced by a factor 0.75 (see text). Middle: The experimental data and the predicted curves for $V$=-25, -50, -75, -100 and -150 MeV divided by the calculation for scenario of $V$=0 MeV and presented on a linear scale. The colour code is identical to the one in Fig.\ref{fig:exc}. Right: $\chi^{2}$-fit of the data with the calculated momentum distributions for the different scenarios.}
\label{fig:mom}
\end{figure}

\subsection{Momentum distribution of the $\eta'$ mesons}
\label{sec:mom}
As a consistency check for the deduced $\eta^\prime$-potential depth the momentum distribution of $\eta^\prime$ mesons, which is also sensitive to the potential depth, has been investigated as well. A comparison of the measured and calculated momentum distributions in the incident photon energy range 1500-2200 MeV is shown in Fig.~\ref {fig:mom} left. The momentum resolution varies between 25-50 MeV/c deduced from the experimental energy resolution and from MC simulations and is smaller than the chosen bin size of 100 MeV/c. In Fig.~\ref{fig:mom} middle the experimental data and the scenarios with potential depths V=-25, -50, -75, -100 and -150 MeV are divided by the calculation for V=0 MeV and are shown on a linear scale. The comparison of data and calculations again seems to exclude strong $\eta^\prime$ mass shifts. A $\chi^{2}$-fit of the data with the calculated momentum distributions for the different scenarios (see Fig.~\ref{fig:mom} right) over the full range of incident energies gives a potential depth of -(32$\pm$11) MeV.\\

\par
The difference in deduced values for the potential depth reflects the systematic uncertainties of the present analysis. With proper weighting of the errors an over all value of V$_0 (\rho = \rho_0) = -(37 \pm 10(stat)\pm10(syst))$ MeV is deduced.\\

\section{Conclusions}
 Experimental approaches to determine the $\eta^\prime$-nucleus optical potential have been presented and discussed. The imaginary part of the $\eta^\prime$-nucleus optical potential, deduced from transparency ratio measurements, has been found to be (-10$\pm$2.5) MeV \cite{nanova}.
Within the model used, the present results on the real part of the potential are consistent with an attractive $\eta^\prime$-nucleus potential with a depth of -($37 \pm 10(stat)\pm10(syst)$) MeV. This result implies the first (indirect) observation of a mass reduction of a pseudo-scalar meson in a strongly interacting environment under normal conditions ($\rho=\rho_{0}, T=0$). The attractive $\eta^\prime$-nucleus potential might even be strong enough to allow the formation of bound $\eta^\prime$-nucleus states. The search for such states is encouraged by the relatively small in-medium width of the $\eta^\prime$~\cite{nanova}. Experiments are proposed to search for $\eta^\prime$ bound states via missing mass spectroscopy~\cite{Kenta} at the Fragment Separator (FRS) at GSI and in a semi-exclusive measurement at the BGO-Open Dipol (OD) setup at the ELSA accelerator in Bonn~\cite{volker}, where observing the formation of the $\eta^\prime$-mesic state via missing mass spectroscopy will be combined with the detection of its decay. A corresponding semi-exclusive experiment is also proposed for the Super-FRS at FAIR~\cite{Nagahiro_Kenta}. The observation of $\eta^\prime$-nucleus bound states would provide further direct information on the in-medium properties of the $\eta^\prime$ meson.
  \\

\section{Acknowledgements}
We thank the scientific and technical staff at ELSA and the collaborating
institutions for their important contribution to the success of the
experiment. We acknowledge detailed discussions with U. Mosel and K. Itahashi. This work was supported financially by the {\it
 Deutsche Forschungsgemeinschaft} within SFB/TR16 and by the {\it Schweize\-ri\-scher Nationalfonds} .

\end{document}